\documentclass[aps,prd,onecolumn,superscriptaddress,showpacs]{revtex4}

\usepackage{graphicx}
\usepackage{dcolumn}
\usepackage{bm}
\usepackage{pstricks}
\usepackage{color}

\newcommand{\be}{\begin{equation}}
\newcommand{\ee}{\end{equation}}
\newcommand{\bea}{\begin{eqnarray}}
\newcommand{\eea}{\end{eqnarray}}

\newcommand{\lsim}{ \mathop{}_{\textstyle \sim}^{\textstyle <}}
\newcommand{\vev}[1]{ \left\langle {#1} \right\rangle }

\newcommand{\Mpc}{{\rm ~Mpc}}

\newcommand{\s}{{\rm ~s}}

\newcommand{\cm}{{\rm ~cm}}
\newcommand{\km}{{\rm ~km}}

\newcommand{\K}{{\rm ~K}}
\newcommand{\mK}{\rm ~mK}

\newcommand{\keV}{{\rm ~keV}}
\newcommand{\MeV}{{\rm ~MeV}}
\newcommand{\GeV}{{\rm ~GeV}}
\newcommand{\TeV}{{\rm ~TeV}}
\newcommand{\kpb}{{\rm ~keV/baryon}}

\newcommand{\epp}{e^\pm}

\begin{document}

\title{CMB and 21-cm Signals for Dark Matter with a Long-Lived Excited State}

\author{Douglas P. Finkbeiner}
\affiliation{Harvard-Smithsonian Center for Astrophysics, 60 Garden St., Cambridge, MA 02138}

\author{Nikhil Padmanabhan}
\affiliation{Physics Division, Lawrence Berkeley National Laboratory, 1 Cyclotron Rd., Berkeley, CA 94720}

\author{Neal Weiner}
\affiliation{Center for Cosmology and Particle Physics, Department of Physics, New York University, 
New York, NY 10003}

\date{\today}

\begin{abstract}

Motivated by the eXciting Dark Matter (XDM) model of Finkbeiner \& Weiner, 
hypothesized to explain the 511 keV signal in the center of the Milky Way,
we consider the CMB and 21-cm signatures of models of dark matter
with collisional long-lived excited states. 
We compute the relic excitation fraction
from the early universe for a variety of assumptions about the
collisional de-excitation cross-section and thermal decoupling. The
relic excitation fraction can be as high as 1\% for natural regions
of parameter space, but could be orders of magnitude smaller. Since
the lifetime of the excited state is naturally greater than $10^{13}$s, we
discuss the signatures of such relic excitation on cosmic microwave
background (CMB) and high-$z$ 21-cm observations. Such models have 
potentially richer astrophysical signals than the traditional WIMP
annihilations and decays, and may have observable consequences for 
future generations of experiments.

\end{abstract}


\maketitle
\twocolumngrid

\section{Introduction}

To explain the apparent excess of $\epp$ annihilation in the
Galactic bulge observed by the INTErnational Gamma-Ray Astrophysics
Laboratory (INTEGRAL)
\cite{Weidenspointner:2006nu, Weidenspointner:2007},
Finkbeiner \& Weiner \cite{Finkbeiner:2007kk} proposed a model of
eXciting Dark Matter (XDM) in which Weakly Interacting Massive
Particles (WIMPs) collisionally excite and subsequently de-excite via
$\epp$ emission.  This model uses the kinetic energy of the WIMP dark
matter to create $\epp$ pairs, in contrast with light dark matter
models in which the pairs result from the mass energy of WIMP
annihilation \cite[e.g.,][]{Boehm:2003bt}, where the WIMP mass must be
less than a few MeV \cite{Beacom:2004pe, Beacom:2005qv}.  Because the
XDM WIMP must have a weak-scale mass ($\sim 500 \GeV$) it retains many
of the desirable properties of weak-scale WIMPs such as the thermal
relic freeze-out abundance.

The lifetime of such an excited state need not be short, and indeed, could
be of order the age of the universe today. This raises the possibility of
a long-lived relic excited fraction with observable consequences.
A simple argument shows the large amount of energy potentially
available from de-excitations - assuming 
100\% of the DM is the XDM WIMP, and the relic
excitation fraction is $Y_f$, the energy per
baryon, $p$, is
\be
p = Y_f \eta M_\chi \frac{n_\chi}{n_b}
  = Y_f \eta \frac{\rho_{DM}}{\rho_b} m_p
\ee
where $\eta$ is the fraction of the WIMP mass converted to
kinetic energy by the
de-excitation, $M_\chi$ is the WIMP mass, and 
$\rho_{DM}/\rho_{b} \approx 5$.  For the fiducial
XDM model, we take
\be
\eta = (\delta - 2m_e)/M_\chi,
\ee
which for mass splitting $\delta \approx 1.1-2$ MeV, and $M_\chi =
500$ GeV yields $\eta \approx 2\times 10^{-7} $ to $ 2\times 10^{-6}$,
or
\be
\label{eq:evperbaryon}
p \approx Y_f (1-10\kpb)
\ee
This amount of energy, even if inefficiently transferred to the gas,
could completely ionize the universe many times over for $Y_f=1/2$.  For the
more realistic case of $Y_f \ll 1/2$, the consequences depend on
when and where the energy is deposited, and with what efficiency.

This paper explores the astrophysical phenomenology of XDM WIMP relic
excitations. We start by showing that for a natural range of
cross-sections, the residual excited fraction can be high enough ($>
10^{-4}$) to have measurable consequences. We then explore these
consequences, focusing on the ionization and thermal history of the
universe, and discuss how observations of the cosmic microwave
background and diffuse 21-cm radiation might constrain such effects.
Our goals here are two-fold: to determine whether the specific model
of XDM proposed to explain the 511 keV excess is constrained by other
astrophysical probes, and to explore more generally the phenomenology
of a WIMP with one or more excited states.  As we shall show, this
more general class of ``XDM''-like models could have a much richer
astrophysical phenomenology than traditional WIMPs.

\section{Kinetic Decoupling and Decays of XDM Particles}
Before addressing the implications of excited states on reionization, we must address two questions within the
context of the model: how does the kinetic temperature of the XDM relate to the photon
temperature when de-excitation goes out of equilibrium, and what is the lifetime of the excited state $\chi^*$\,?
The former question is important for determining the precise value of the relic density of $\chi^*$, while the
latter is important for the transfer of energy from the $\chi^*$ to
ionization in the later universe.

\subsection{Summary of the XDM model}
The defining feature of the XDM model is that the WIMP has an excited
state which can be collisionally excited, and subsequently decay to
$e^+e^-$ pairs.  The excited state could exist due to compositeness of
the dark matter, or arise from an approximate symmetry of the theory.

For the excited state to be accessible in the Milky Way, and relevant
for $e^+e^-$ production, only a narrow kinematical range must be
considered for the mass splitting, $\delta$.  For the decay to the
ground state to be energetically capable of producing $e^+e^-$ pairs,
one must have $\delta > 1.022 \MeV$. On the other hand, the kinetic
energy available for a pair of $500 \GeV$ WIMPs colliding each with
velocities $v \sim 600 \km/\s$ (roughly the escape velocity of the
galaxy), is $2\MeV$, setting an upper bound on $\delta$.

To produce a sufficiently high number of positrons to explain the
INTEGRAL signal, a large cross section is required
\cite{Finkbeiner:2007kk,Pospelov:2007xh}, comparable to the geometric
cross section set by the characteristic momentum transfer. That is,
$\sigma \sim (M_\chi \delta)^{-1}$ is of the correct size. Such a
cross section can arise naturally \cite{Finkbeiner:2007kk}, but
requires the presence of a new light scalar $\phi$, with $m^2_\phi
\lsim M_\chi \delta$. The $\chi$ can excite by emitting a $\phi$ with
amplitude $\lambda_-$ or can scatter elastically with amplitude
$\lambda_+$. We generally assume $\lambda_-\sim \lambda_+$, but this
isn't necessary.

Most of the equilibrium properties relevant to our discussion here are ultimately set by the interactions of $\phi$, which stays in thermal
equilibrium with the standard model through its mixing with the Higgs. Thus, the most relevant term for the
discussion at hand is the $\phi$-Higgs coupling \be {\cal L} \supset \alpha \phi^2 h^\dagger h. \ee

When the Higgs acquires a vacuum expectation value (vev), this contributes to the mass of the $\phi$. Thus
requiring a tuning better than $1\%$ in parameters yields a naturalness upper bound of about $\alpha \lsim 10^2 \, m_\phi^2/v^2 \sim 2 \times 10^{-3} (m_\phi^2/1 \GeV^2)$. Assuming a vev, $\vev{\phi} \sim m_\phi$, one finds a mixing angle between the $\phi$ and Higgs
of $\sin \theta \approx \alpha\, m_\phi v/m_h^2 \lsim 10^{-4}\,
$\footnote{
It is possible that the vev of $\phi$ could be smaller than $m_\phi$
if there are other sources of symmetry break generating the splitting
between $\chi$ and $\chi^*$. Such effects could even be beneficial for
early universe cosmology \cite{Finkbeiner:2007kk}. If such effects are
present, they could also contribute to the mixing of $\phi$, for
instance through terms such as $A \phi h^\dagger h$. Nonetheless, such
terms should be related to $m_\phi$ through radiative corrections, and
we expect the scaling of the mixing angle with $m_\phi$ to be fairly
robust in a wide variety of theories.}.
Note that the natural range of mixing angle is correlated with $m_\phi$. That is, since $\alpha \lsim 10^2 \, m_\phi^2/v^2$, $\sin \theta \lsim 10^2 m_\phi^3/(100 \GeV)^3$. Thus, very light $\phi$'s are naturally more weakly mixed than heavier $\phi$'s.

\subsection{Kinetic Decoupling of XDM} Although XDM annihilation $\chi \chi \leftrightarrow \phi \phi$ freezes
out in a fashion similar to usual WIMPs at $T\sim M_\chi/20$, kinetic decoupling is a somewhat more subtle story.
Direct elastic scattering $\chi f \rightarrow \chi f$ is both Yukawa and mixing suppressed, and is thus
inefficient at maintaining kinetic equilibrium.

\begin{figure}
\begin{center}
\leavevmode
\includegraphics[width=1.5in]{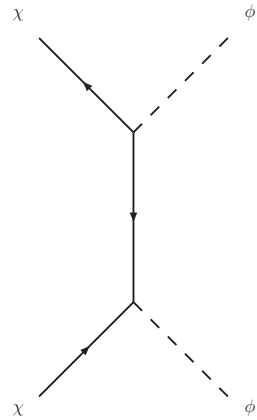}
\end{center}
\caption{Dominant diagram contributing to kinetic equilibrium of $\chi$.}
\label{fig:kineq}
\end{figure}

The dominant process contributing to kinetic equilibrium of $\chi$ is $\chi \phi \rightarrow \chi \phi$ shown in
Fig. \ref{fig:kineq}. The scattering cross section for this process is
\be
\sigma = \frac{\lambda^4}{4 \pi m_\phi^2},
\ee
where $\lambda$ is the $\chi - \chi -\phi$ coupling. (We assume $\lambda_+ \sim  \lambda_-$ here for simplicity, although that does not significantly change this discussion. Additionally, we do not distinguish between $\chi$ and $\chi^*$ at this temperature $T \gg \delta$.) 
\begin{figure}
\begin{center}
\leavevmode
\includegraphics[width=1.5in]{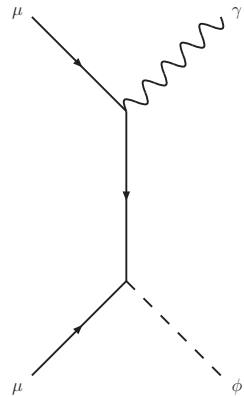}
\end{center}
\caption{Dominant diagram contributing to thermal equilibrium of $\phi$.}
\label{fig:phieq}
\end{figure}
With such a cross section and $\lambda \sim 0.1$, assuming a thermal presence of $\phi$, $\chi$ will remain in
kinetic equilibrium down to $T \simeq m_\phi/30$.

Ultimately, the relevant process for determining the decoupling temperature is when $\phi$ decouples from the
thermal bath. The dominant diagram for this process is shown in Fig. \ref{fig:phieq}. Since we are principally
interested in the lowest possible temperature $T_{dec}$, we are interested in the situation where the fermion in
question is a muon, and the cross section (for relativistic $\mu$ and non-relativistic $\phi$) is approximately
given by
\be
\sigma = \frac{y_\mu^2 \sin^2 \theta \alpha_{\rm em}}{8 \pi m_\phi^2},
\ee
where $\alpha_{\rm em}$ is the fine structure constant.

Such a scattering can keep the $\phi$ in equilibrium down to below the muon mass for $\sin \theta = 10^{-4}$,
while for smaller mixings the temperature of decoupling is higher (roughly 1 GeV for $\sin \theta = 10^{-5}$,
where additional fields, such as pions and kaons, are relevant). As such, we limit ourselves to the
range of $T_{dec} > 100$ MeV, although one could conceivably stay in equilibrium longer in other
models.

\subsection{Lifetime}
In addition to the couplings that drive the early thermal history, the
lifetime of both $\chi^*$ and $\phi$ are clearly important. The
lifetime of $\chi^*$ is crucial, because this determines when the
energy of the excited state can be deposited into the baryonic gas of
the early universe. The lifetime of $\phi$ is important, as we have
ignored its presence in the calculations at $T \approx 1 {\rm MeV}$
and we need to see that this is justified.

We begin by considering the lifetime of the $\chi^*$. Both excited and unexcited states of the dark matter will
come into thermal equilibrium in the early universe, with the excited state decaying into the lighter state with
an approximate lifetime
\be 
\tau_{\chi^*} \approx 10^{15}{\rm s}
\left(\frac{ 0.1}{\lambda_-}\right)^2
\left(\frac{10^{-4}}{\sin \theta}\right)^2 
\left(\frac{1\rm MeV}{\sqrt{\delta^2-4 m_e^2}}\right)^5 
\left(\frac{ m_\phi}{1 \rm GeV}\right)^4. 
\ee 
Consequently, for the parameters under consideration, lifetimes in the range of $10^{13}$s to $10^{18}$s
are quite reasonable.

Although it appears one can make the lifetime much longer simply by
lowering $m_\phi$, there is an implicit link between $m_\phi$ and
$\sin \theta$ because of naturalness. Lower values of $m_\phi$ may
quickly lead to a highly tuned region of parameter space.

The scalar $\phi$ decays through its mixing with the Higgs, and thus has a lifetime 
\be 
\tau_\phi = \sin^{-2}
\theta\, \tau_h(m_h = m_\phi). 
\ee 
That is, the lifetime of $\phi$ is just $\sin^{-2}\theta$ times that
of a Higgs boson with mass $m_\phi$. This decay is typically dominated
by a single process. For example, for $m_\phi \lsim 1 \MeV$, the
$\phi$ decays into $\gamma \gamma$. A Higgs of this mass has a
lifetime $\tau \sim 3 \times 10^{-4} {\rm s}$
\cite{Gunion:1989we}. Thus, the $\phi$ will have a lifetime
$\tau_\phi \sim 3\sin^{-2} \theta \times 10^{-4} {\rm s}$. Although
such extremely light $\phi$ bosons are potentially interesting, they
occupy a very tuned region of parameter space.

For $2 m_e \lsim m_\phi \lsim 2 m_\mu$, $\phi \rightarrow e^+ e^-$
dominates, and the $\phi$ lifetime will range $10^{-9} \sin^{-2}
\theta \lsim \tau_\phi/(1{\rm~s}) \lsim 3 \times 10^{-11} \sin^{-2}
\theta$. Thus, for mixing angles $\sin \theta \sim 10^{-4}$,
such particles would decay before nucleosynthesis, and in general well
before kinetic decoupling of $\chi$.  The most natural region of
parameter space (with the lowest tuning), $2 m_\mu \lsim m_\phi \lsim
1 \GeV$, $\phi \rightarrow \mu^+ \mu^-$ and $\phi \rightarrow \pi \pi$
become available, and the $\phi$ lifetime will range $3 \times
10^{-15} \sin^{-2} \theta \lsim \tau_\phi/(1{\rm~s}) \lsim 3 \times
10^{-18} \sin^{-2} \theta$.  In this range of parameters,
$\phi$ will have certainly decayed before kinetic decoupling of
$\chi$.

Thus, with reasonable values for the mixing parameter, we find the
scalar is relatively short-lived. However, the decay of the excited
state $\chi^*$ into the lighter state $\chi$ will naturally occur late
in the universe, producing positrons and feeding energy into the
baryonic fluid. The amount of energy will depend directly of the
number of relic $\chi^*$'s left over from the Big Bang.

\section{Collisional XDM Freeze-out}

Excitation and de-excitation of the WIMP proceeds at rates $k_{E}$ and $k_{D}$, respectively, via reactions of the form
\begin{eqnarray}
\label{eq:react}
\chi \chi & \leftrightarrow & \chi^{*} \chi \\
\label{eq:reactb}
\chi \chi^{*} & \leftrightarrow & \chi^{*} \chi^{*} \,\,.
\end{eqnarray}
For simplicity, we assume that the rates for both channels
(\ref{eq:react},\ref{eq:reactb}) are equal, and neglect double
excitations and double de-excitations.  We denote the physical (not
comoving) densities of $\chi, \chi^{*}$ by $n_{\chi}, n_{\chi^{*}}$
and define $n \equiv n_{\chi} + n_{\chi^{*}}$. The Boltzmann equation
for $n_{\chi^{*}}$ is then
\begin{equation}
\label{eq:boltznchis}
\frac{dn_{\chi^{*}}}{dt} + 3H(t)n_{\chi^{*}} = -k_{D}n_{\chi^{*}}n + k_{E}n_{\chi}n \,\,,
\end{equation}
where $H(t)$ is the Hubble constant at time $t$.
Defining $Y \equiv n_{\chi^{*}}/n$ and a dimensionless inverse temperature $x \equiv \delta/T$,
this simplifies to 
\begin{equation}
\label{eq:boltzY}
\frac{dY}{dt} = -k_{D}n \left[Y - (1-Y)f(x)e^{-x}\right] \,\,.
\end{equation}
where 
\begin{equation}
f(x) = \sqrt{1+\frac{\pi x}{4}} \,\,,
\end{equation}
which is derived for the relation between the excitation and
de-excitation rates for a suitable approximation of the cross section
(see Appendix~\ref{sec:rates}).

\begin{figure}
\begin{center}
\leavevmode
\includegraphics[width=3.0in]{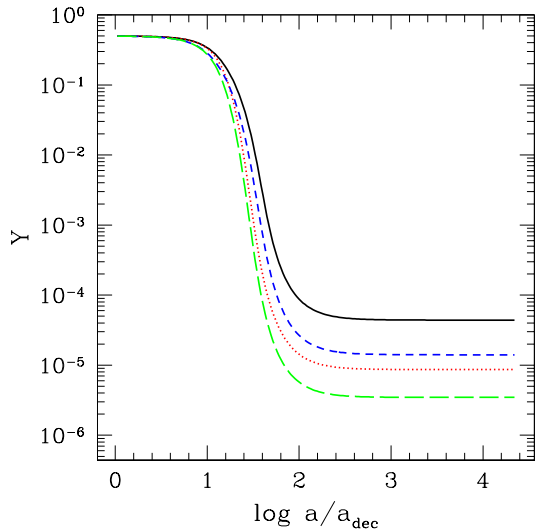}
\end{center}
\caption{The evolution of the excitation fraction with time, for
$M_\chi = 1\,{\rm TeV}$, $T_{\rm dec}=100\,{\rm MeV}$, $\delta=1\,{\rm
MeV}$ and $\tilde{\sigma}_{mr}=1$ (see
Eq. \ref{eq:app_kd_fid}). Solutions are shown for the full coupled
Boltzmann equations (\emph{solid black}). We show the results under certain additional approximations as well, in particular with
temperature coupling turned off (\emph{dashed blue}), for the
approximation $k_{E} = k_{D} \exp(-\delta/T)$, ignoring the
corrections in Appendix~\ref{sec:rates} (\emph{red dotted}), and for
temperature coupling turned off \emph{and} $k_{E} = k_{D} \exp(-\delta/T)$
(\emph{long-dashed green}).
}
\label{fig:yhistory}
\end{figure}

We assume that the $\chi$ particles have decoupled from the radiation at a much earlier time than is 
relevant for the freeze-out of the excited states, i.e. $T_{\rm dec} \gg \delta$;
the kinetic temperature $T$ and the photon temperature $T_{\gamma}$ are then related by
$T_{\gamma} \approx \sqrt{T_{\rm dec} T}$.
The kinetic temperature evolves as
\begin{equation}
\label{eq:boltzT}
\frac{dT}{dt} = - 2H(t)T - \frac{2\delta}{3} \left[k_{E}n_{\chi} - k_{D}n_{\chi^{*}}\right] \,\,.
\end{equation}
The first term describes the adiabatic cooling of the WIMPs, while the second 
is the thermal energy absorbed and injected as the WIMPs excite and de-excite.
This implicitly assumes that the elastic scattering cross-section is much greater than 
the excitation/de-excitation cross-sections; this ensures that the kinetic energy gained/lost
through de-excitations/excitations is efficiently thermalized and the 
$\chi$ particles maintain a Boltzmann distribution.
Comparing the RHS of Eq.~\ref{eq:boltznchis}
and Eq.~\ref{eq:boltzT}, it is possible to simplify Eq.~\ref{eq:boltzT} and obtain
\begin{equation}
\label{eq:boltzT2}
\frac{d(x^{-1})}{dt} + \frac{2H(t)}{x} = -\frac{2}{3}\frac{dY}{dt} \,\,.
\end{equation}
Substituting $\ln(\tilde{a}) \equiv \ln(a/a_{\rm dec})$ for the time variable, where
$a_{\rm dec}$ is the scale factor at kinetic decoupling, yields the following 
coupled Boltzmann equations:
\begin{equation}
\label{eq:boltzT3}
\frac{dx}{d\ln \tilde{a}} = 2x + \frac{2x^2}{3}\frac{dY}{d\ln \tilde{a}} \,\,,
\end{equation}
and from Eq.~\ref{eq:boltzY},
\begin{equation}
\label{eq:boltznchis2}
\frac{dY}{d\ln \tilde{a}} = - \frac{\alpha_{\rm dec}}{4\tilde{a}} \sqrt{\frac{x_{dec}}{x}}
\left[Y - (1-Y)f(x)e^{-x}\right] \,\,.
\end{equation}
where $\alpha_{\rm dec} \equiv k_{D}(a_{\rm dec}) n(a_{\rm dec})/H(a_{\rm dec})$. Using
our fiducial cosmology, we estimate $\alpha_{\rm dec}$
\begin{eqnarray}
\label{eq:alpha_est}
&& \alpha_{\rm dec} \sim 10^{8} \tilde{\sigma}_{mr}  \nonumber \\
&&\times \left(\frac{T_{\rm dec}}{1 {\rm GeV}}\right)^{3/2}
\left(\frac{M_\chi}{100 {\rm GeV}}\right)^{-5/2} \left(\frac{\delta}{1{\rm MeV}}\right)^{-1}.
\end{eqnarray}

\begin{figure}
\begin{center}
\leavevmode
\includegraphics[width=3.0in]{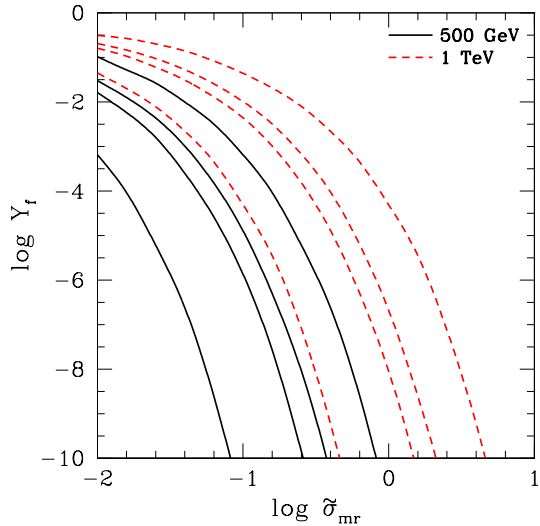}
\end{center}
\caption{The residual excitation fraction $Y_f$ as a function of the
scattering cross section $\tilde{\sigma}_{mr}$, and decoupling
temperature $T_{\rm dec}$, for $M_\chi = 500\GeV$ (\emph{solid black})
and $M_\chi = 1\TeV$ (\emph{dashed red}). From top to bottom for each
mass, the decoupling temperatures are $T_{\rm dec}$ = 100 MeV, 500
MeV, 1 GeV, and 10 GeV. The 511 keV signal favors $\tilde{\sigma}_{mr} = 0.1$ - $50$
for a 500 GeV WIMP.}
\label{fig:yinf}
\end{figure}

The evolution of the excited fraction with time
(Fig.~\ref{fig:yhistory}) exhibits the expected features: $Y = 1/2$
until the temperature reaches $\sim \delta$, after which time it
rapidly falls until the Hubble expansion shuts off the de-excitation
reactions and it asymptotes to its freeze-out value, $Y_{f}$.  We also
observe that the modifications to the simplest formulation - 
the difference between $k_{E}$ and $k_{D}$, and the change
in the gas temperature due to the changing fraction of $\chi$ in the
excited state - have comparable effects on the freeze-out abundance.
Plotting $Y_{f}$ as a function of various XDM parameters
(Fig.~\ref{fig:yinf}), we find that a significant residual fraction
can survive in some cases, though $Y_f$ is smaller for parameters
favored by the INTEGRAL signal \cite{Weidenspointner:2006nu,
Weidenspointner:2007}, as computed in \cite{Finkbeiner:2007kk}.

While it is convenient to use $\alpha_{\rm dec}$ and $x_{\rm dec}$ to define the initial conditions,
there is an alternative which yields a useful scaling of $Y_{f}$ with $\alpha_{\rm dec}$ and $x_{\rm dec}$.
To derive this, we start by noting that the combination $c = (\alpha_{\rm dec}/\tilde{a})\sqrt{x_{\rm dec}/x}$
(which compares the de-excitation rate to the Hubble expansion) controls the behavior of the system. While $c \gg 1$,
$Y$ remains at its equilibrium value, and the kinetic temperature simply evolves as $a^{-2}$. This suggests
defining the initial condition as the epoch when $c = c_{0} \gg 1$.  This occurs when 
$\tilde{a}^{2} = \alpha_{\rm dec}/c_{0}$, or when $x_{0} = x_{\rm dec} \alpha_{\rm dec}/c_{0}$. If $x_{0} \ll 1$,
then Eqs.~\ref{eq:boltzT3} and \ref{eq:boltznchis2} will have identical initial starting points if 
$\alpha_{\rm dec} x_{\rm dec}$ is the same, as $c_{0}$
is just an arbitrary constant. This implies that 
the freeze-out value only depends on the combination $\alpha_{\rm dec} x_{\rm dec}$.
This will not be true if $x_{0} > 1$, but in this case, the system 
will remain in equilibrium as the temperature falls below $\delta$ and the residual fraction
will be exponentially suppressed to an uninteresting value. 
The combination $\alpha_{\rm dec} x_{\rm dec}$ scales with the XDM parameters as 
\begin{equation}
\label{eq:alphax}
\alpha_{\rm dec} x_{\rm dec} = 10^{5} \tilde{\sigma}_{mr} 
\left(\frac{T_{\rm dec}}{1 {\rm GeV}}\right)^{1/2} \left(\frac{M_\chi}{100 {\rm GeV}}\right)^{-5/2} \,\,.
\end{equation}
Interestingly, it is independent of $\delta$, at least for the
assumption of $T_{\rm dec} \gg \delta$ made above.  This relationship
provides a better understanding of the scaling of the curves in
Fig.~\ref{fig:yinf}.  In spite of the exponential uncertainty in the
relic excitation, it is natural to have significant and interesting relic excitation fractions. Thus, we now turn to the observable consequences of such a large $Y_f$.

\section{Observational Signatures}
\label{sec:observe}
In this section we consider the consequences of WIMP de-excitation at
early enough times ($z>10$) that the collisional excitation and
de-excitation expected for centers of halos at late times is
unimportant.

The most obvious observables affected by energy injection into the IGM
during the ``dark ages'' ($10 < z < 500$) are the CMB and the high-$z$
21 cm background.  The effect of DM annihilations and decays on the CMB has
previously been discussed 
\cite{Peebles:2000pn,Padmanabhan:2005es,Mapelli:2006ej};
the case at hand is more like DM decay. The energy injection from WIMP
decay, whether from sterile neutrinos \cite{Hansen:2003yj}, superheavy
dark matter \cite{Doroshkevich:2002ff}, or generically
\cite{Bean:2003kd}, has been shown to affect the ionization history
\cite{Pierpaoli:2003rz}, though if the lifetime of the decaying
particle is longer than the age of the universe, perturbations to the
CMB power spectrum are small \cite{Chen:2003gz,Mapelli:2006ej}.
Effects on the high-$z$ 21 cm \cite{Furlanetto:2006wp,Valdes:2007cu}
and structure formation \cite{Ripamonti:2007} have also been
investigated. As we shall see, the XDM WIMP exhibits a much larger
range of observable signals for natural models, in part because 
a potentially large fraction of the DM can participate, but also
because the characteristic energy scales are significantly lower and 
can be absorbed efficiently by the gas.

We begin by estimating the relevant time scales for energy deposition,
and demonstrate that the $e^{\pm}$ pairs produced by XDM de-excitations 
deposit their kinetic energy efficiently. We then consider the possible
consequences of this energy deposition and the potential for detecting it in 
current and future experiments.

\subsection{Energy Deposition Timescales}
\label{sec:times}

The de-excitation of $\chi^{*}$ deposits energy into the intergalactic medium
in the form of non-relativistic electron-positron pairs with kinetic energies $\sim 100\keV$.
The dominant energy loss for such electrons is via collisions, with a
cross-section for collisional ionization \cite{Shull:1985} of
\begin{equation}
\label{eq:collcrosssection}
\sigma_{eH} = \frac{2.23 \times 10^{-15} \ln(E/13.6)}{E}\cm^2 \,\,,
\end{equation}
where $E$ is the kinetic energy in eV.  The cross-sections for
excitations and heating are similar. For a 100 keV electron, this
corresponds to a cross-section of $\sim 2 \times 10^{-19}\cm^2$, 
or a scattering rate of
\begin{equation}
n_{H} \sigma_{eH} v \sim 5 \times 10^{-13} \left(\frac{1+z}{10}\right)^{3} {\,\rm s}^{-1} \,\,.
\end{equation}
Comparing this to the Hubble time 
\begin{equation}
\frac{1}{H(z)} \sim 10^{16}h^{-1} \left(\frac{1+z}{10}\right)^{-3/2} {\rm s}
\end{equation}
we see that collisional energy deposition is extremely efficient over the 
entire redshift range of interest. We assume
that all the kinetic energy of the electrons is instantaneously partitioned between 
ionizations, heating and excitations.

The above has focused on the deposition of the kinetic energy of the
$e^{\pm}$ pairs; there is an additional $\sim 1\MeV$ available from
the rest mass energy of $e^{\pm}$.  Positrons can annihilate to 2
photons at 511 keV, or to 3 photons.  The $3\gamma$ spectrum of
ortho-positronium \cite{Ore:1949te} is very hard, with only $7\times
10^{-3}$ of the power coming out at $E_\gamma < 100\keV$ and $7\times
10^{-6}$ at $E_\gamma < 10\keV$.  At redshift $z < 100$, the universe
is nearly transparent to these photons, so their energy is effectively
lost.  At $z>100$ the photon energy density is high enough that
Compton scattering happens faster than a Hubble time, so for relevant
lifetimes ($\tau_{\chi^{*}} \approx 10^{13} - 10^{14}\s$) the mass
energy of the pair must be included \cite[see
  e.g.][]{Ripamonti:2006gq}, giving rise to a higher effective $\eta$.

\subsection{Ionization/Thermal History}

The effects of XDM on the ionization and thermal history of the Universe are 
controlled by two parameters - the available energy per baryon $\epsilon_{b}$
and the lifetime of the excited state $\tau_{\chi^{*}}$. The energy per 
baryon is determined both by the energy splitting $\delta$ and the residual
excitation fraction $Y_{f}$,
\begin{equation}
\label{eq:eperbaryon}
\epsilon_{b} = Y_{f} \frac{n_{\chi}}{n_{b}} \left(\delta - 2m_{e}c^{2}\right) \,\,,
\end{equation}
where the $n_{b,\chi}$ are the number densities of baryons and XDM particles.
To more easily connect to the relic abundance calculation of the
previous section, we fix $\delta - 2m_{e}c^{2}$ to 100 keV and $n_{\chi}/n_{b}$
to $10^{-2}$ below, corresponding to a WIMP mass of 500 GeV. 
The energy per baryon is then trivially related to the
relic abundance by 
\begin{equation}
\epsilon_{b} \approx 10^{3} Y_{f} {\,\,\rm eV} \,\,,
\end{equation}
allowing us to express our results in terms of $Y_{f}$ and $\tau_{\chi^{*}}$.
It is straightforward to relate the results below to cases which make
different assumptions for the energy splitting and number densities.  

\begin{figure}
\includegraphics[width=3in]{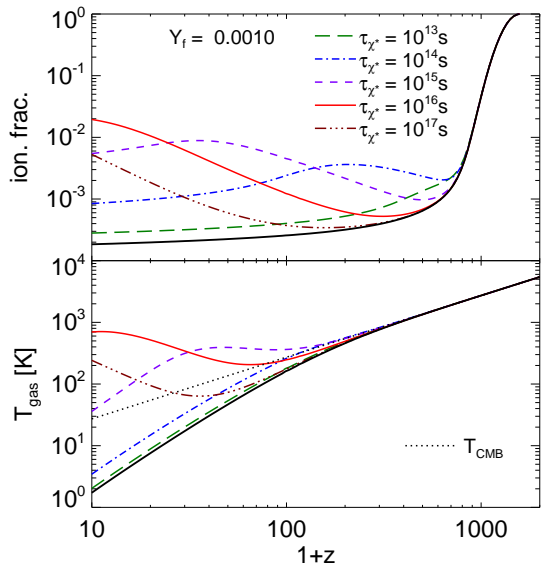}
\caption{\label{fig:igm_temp_yf} The ionization fraction $x_{i}\equiv
 n(e^-)/n(H)$ (\emph{top}), and matter temperature (\emph{bottom}), for
 various values of the lifetime, $\tau_{\chi^*}$, with $Y_f$ held
 fixed.  The baseline scenario, with no energy injection from WIMPs,
 is shown in both panels (\emph{thick solid line}), and $T_{cmb}$ is
 included in the bottom panel (\emph{dotted line}).  In all cases we
 take $Y_f = 10^{-3} \epsilon_{b}$. Note that we ignore the effects 
 of star formation etc. on the ionization fraction and temperature.
}
\end{figure}

\begin{figure}
\includegraphics[width=3in]{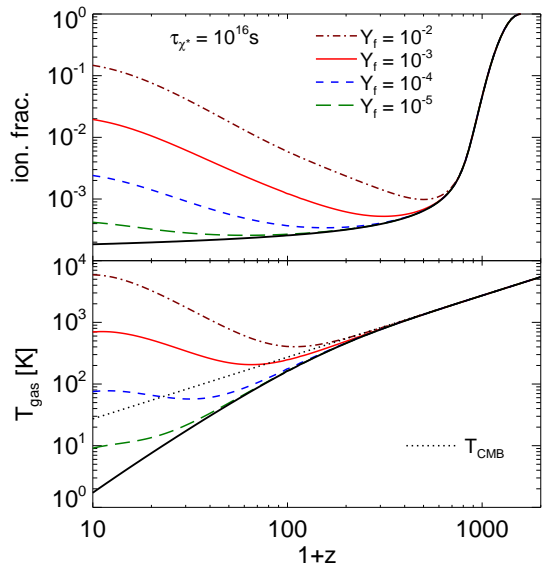}
\caption{\label{fig:igm_temp_tau} Same as Fig. \ref{fig:igm_temp_yf},
 but for
 various values of $Y_f$ with $\tau_{\chi^*} = 10^{16}$s.  The
 fiducial model ($Y_f=10^{-3}$ and $\tau_{\chi^*} = 10^{16}$s) is
 represented by a red solid line in both figures. 
}
\end{figure}

We modify the publicly available code \texttt{RecFast} \footnote{
\texttt{http://www.astro.ubc.ca/people/scott/recfast.html}} 
to numerically calculate the ionization and thermal histories
\cite{Seager:1999,Seager:2000}.  Examples of these for different choices of XDM
parameters are plotted in Figs.~\ref{fig:igm_temp_yf}
and~\ref{fig:igm_temp_tau}, where we hold one parameter fixed while
varying the other.

The effect of varying $Y_{f}$ at constant lifetime is as one expects, 
with an increasing ionization fraction and temperature as $Y_{f}$,
and therefore $\epsilon_{b}$, increases. The dependence on the lifetime
is more involved. For lifetimes much shorter than the age of the Universe
at recombination ($\tau_{\chi^{*}} \ll 10^{13} {\rm s}$), the energy is simply
injected into a fully ionized medium with no effect. Injecting energy soon
after recombination $\tau_{\chi^{*}} \sim 10^{14} {\rm s}$ can truncate 
recombination early, resulting in a higher residual ionization level. For
even longer lifetimes, 
the injected energy does not perturb the baseline recombination, but partially
reionizes the Universe.
However, unlike the previous 
case, the recombination processes have been shut off by the Hubble expansion,
and so the ionization fraction monotonically increases with time. 
This behavior qualitatively persists for longer lifetimes, but the 
degree of reionization decreases as $\tau_{\chi^{*}}$ becomes a significant
fraction of the age of the Universe today and there simply has not been 
enough time for the $\chi^{*}$ to de-excite.

The gas temperature behaves similarly, except that it remains thermally
locked to the photon temperature until $z \sim 300$, much later than recombination.
Lifetimes shorter than this have little effect on the gas temperature.

The above has concentrated on the homogeneous Universe - the formation
of collapsed halos could modify this in two ways. The first is the
$\chi - \chi^{*}$ collisions could de-excite the $\chi^{*}$. Such
de-excitations only increase the kinetic energy of the colliding WIMPs
and do not inject energy into the IGM. The second effect occurs when
virial motions inside halos can re-excite the dark matter. This only
happens for halos with velocities $> \sqrt{\delta/M_\chi}$ and only
becomes significant after the ionization and temperature have already
been considerably modified by star formation.

\subsection{Implications for CMB and 21-cm Observations}
The only two probes of the $z>10$ ionization and thermal history of
the Universe are the cosmic microwave background (CMB) and the
measurements of the 21-cm hyperfine splitting in hydrogen. While CMB
measurements are now a mature field, they are less constraining
because they are only sensitive to the integrated ionization history
and do not probe the temperature of the IGM. On the other hand, 21-cm
measurements probe both the temperature and ionization as a function
of time, but the experimental techniques are less developed with the
first pathfinder experiments scheduled for the near future. We
consider both of these in turn below.

\subsubsection{CMB power spectrum}
\label{sec:cmb}
The polarization anisotropy of the CMB, and its correlation with the
temperature anisotropy, can provide a powerful constraint on the
ionization history of the universe.  The CMB polarization is
principally induced by Thomson scattering \cite{Rees:1968}, and was
first observed on small scales by the DASI interferometer
\cite{Kovac:2002fg}, and on large scales by the WMAP
\cite{Page:2006hz}.  The small-scale polarization is
sensitive to ionization at the epoch of recombination, while
large-scale measurements probe the epoch of reionization at $z\approx
10$.  Such data are, in principle, sensitive to any perturbation of
the ionization history of the universe caused by new physics, such as
XDM.

\begin{figure}
\begin{center}
\leavevmode
\includegraphics[width=3.0in]{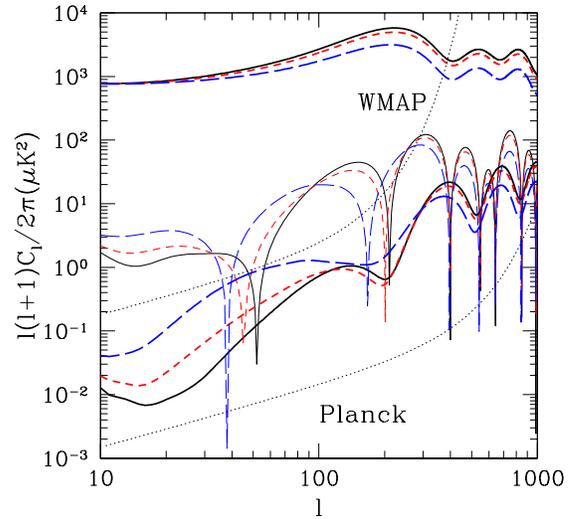}
\end{center}
\caption{The effect of XDM on the CMB TT, TE, and EE power spectra.
The solid (\emph{black}) line is our fiducial cosmology, with no XDM. 
The short-dashed (\emph{red}) line assumes XDM with $Y_{f} = 10^{-3}$,
$M_\chi=500 {\rm GeV}$, $\delta = 1.1 {\rm MeV}$ and $\tau_{\chi^*}=10^{14} {\rm s}$,
while the long-dashed line assumes $Y_{f} = 10^{-2}$ with the other parameters
the same. For lifetimes much less than $10^{13} {\rm s}$, the de-excitation 
of XDM has a negligible effect on the CMB, since the Universe is already completely
ionized. For decays much later, the dominant effect is only on the largest scales,
and therefore hard to disentangle from standard reionization. Also plotted are
nominal curves for the polarization sensitivity in bins of $\log_{10}(\ell)=0.05$ 
for the WMAP and Planck CMB missions.  Uncertainty due to cosmic
variance is not included in these sensitivity estimates. 
}
\label{fig:cmb}
\end{figure}

The effect of XDM on the CMB may be conceptually separated into two
regimes - effects on recombination ($\tau_{\chi^*} \sim 10^{14}{\rm
s}$) and effects on reionization ($\tau_{\chi^*} \sim 10^{16}{\rm
s}$).  The dominant effect on recombination may be thought of as an
increased residual fraction of ionized atoms, which broadens the
surface of last scattering. The increased scattering both washes out
the temperature fluctuations, and enhances and shifts the polarization
power spectra.  These effects on the CMB were discussed in detail by
\cite{Padmanabhan:2005es} for the case of WIMP annihilation, but the
basic physics is also relevant here.
Fig.~\ref{fig:cmb}
plots the temperature and polarization power spectra for two examples
of XDM parameters, with the lifetime chosen to highlight the effects
on recombination. While the differences in the TT power
spectrum are degenerate with the slope of the primordial power
spectrum, these degeneracies are mostly broken by the polarization
power spectra.  Fig.~\ref{fig:cmb} also plots the nominal polarization
sensitivities of current and future CMB measurements.  

The detectability of these changes in the power spectrum will depend
on many factors, including degeneracies with other cosmological
parameters and details of reionization.  Nevertheless, we may estimate
constraints on the $z\approx 1000$ energy injection.  The limit on
such energy injection, marginalizing over the usual cosmological
parameters, is $\epsilon_{DM} < 3\times10^{-14}$ from WMAP, with an
improved limit of
$\epsilon_{DM} < 10^{-15}$ eV/s/baryon at $z=1000$ expected from Planck
\cite{Padmanabhan:2005es}.
Using Eq.~\ref{eq:evperbaryon} for $\eta=2\times10^{-7}$ we have 
\be
\epsilon_{DM} = 10^3 Y_f / \tau_{\chi^*} {\rm ~eV/s/baryon}
\ee
yielding constraints of
\begin{eqnarray}
\label{eq:cmb_constraints}
& Y_f < 3\times10^{-4} (\tau_{\chi^*}/10^{13} \s) & {\rm WMAP}  \nonumber \\
& Y_f < 10^{-5} (\tau_{\chi^*}/10^{13} \s) & {\rm Planck}
\end{eqnarray}
These constraints do not include higher $\ell$ data from e.g. ACBAR
\cite{2008arXiv0801.1491R} and CBI \cite{Sievers:2007}.  Using the
$\epsilon_\alpha$ parametrization for delayed recombination in
(defined by \citet{Peebles:2000pn})
Kim \& Naselsky find that $\epsilon_\alpha < 0.02$ based on WMAP and
ACBAR \cite{Kim:2008}.  This constraint also converts to $Y_f < \sim
2\times10^{-4} (\tau_{\chi^*}/10^{13} \s)$ in agreement with
Eq. \ref{eq:cmb_constraints}.

\begin{figure}
\includegraphics[width=3in]{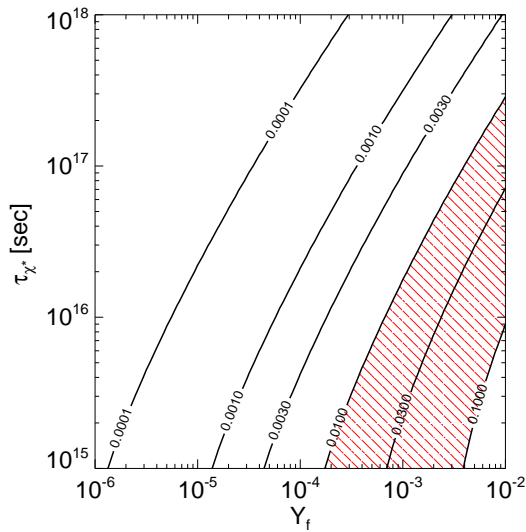}
\caption{\label{fig:cmbtau} The \emph{excess} optical depth for CMB scattering,
  $\Delta\tau$ (see \S \ref{sec:cmb}).  The hatched region is accessible to observation, with
  $\Delta\tau > 0.1$ already ruled out by CMB observations, and $\Delta\tau <
  0.01$ difficult to notice in the presence of the reionization caused
  by standard astrophysics.
}
\end{figure}

\begin{figure}
\includegraphics[width=3in]{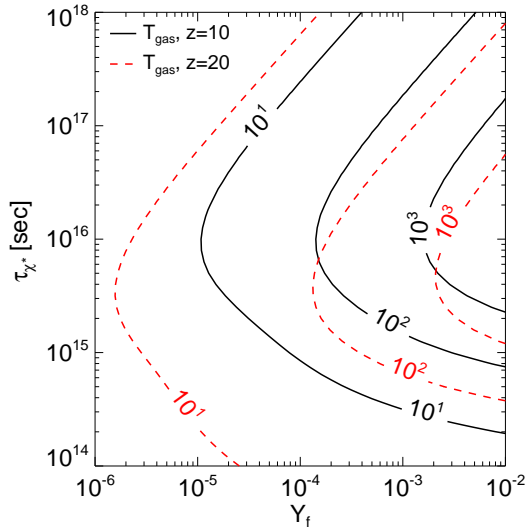}
\caption{\label{fig:gastemp} The gas temperature [K] as a function of 
the freeze-out excitation fraction, $Y_f$, and the lifetime,
$\tau_{\chi^*}$, for two redshifts. 
}
\end{figure}

For longer lifetimes $\tau_{\chi^*} > 10^{15}\s$, the effect on the
reionization history ($z\approx10$) could be pronounced.  In this
limit, it is useful to express the information in the
CMB polarization in the form of a scattering optical depth, $\tau$,
given by
\be
\tau = \frac{n_0\sigma_T c}{H_0} \int_{a_{\rm ref}}^1 \frac{da}{a^4} 
        \frac{x_i(a)}{\sqrt{\Omega_m/a^3 + \Omega_\Lambda}}\,,
\ee
where $n_0$ is the number density of H at $z=0$, $x_i$ is the ionized
fraction of H, $a_{\rm ref}$ is the scale factor at some early
reference time, and the Hubble parameter has been expressed in terms
of $H_0$, $\Omega_m$ and $\Omega_\Lambda$.  We are interested in the
``excess'' optical depth due to $\chi^*$ decay, $\Delta\tau=\tau(Y_f,
\tau_{\chi^*}) - \tau(Y_f=0)$, which we compute for a range of
$Y_f$ and $\tau_{\chi^*}$ (Fig. \ref{fig:cmbtau}).
We identify the region of parameter space where $\Delta\tau$ is
large enough to distinguish from the somewhat uncertain standard
scenario ($\Delta \tau \sim 0.01$) and small enough to be ruled out by
WMAP ($\Delta \tau \sim 0.1$).  While the CMB is a good probe of relic
XDM excitation in some parts of parameter space,
21-cm experiments have a broader reach. 

\subsubsection{21 cm observations}

\begin{figure}
\includegraphics[width=3in]{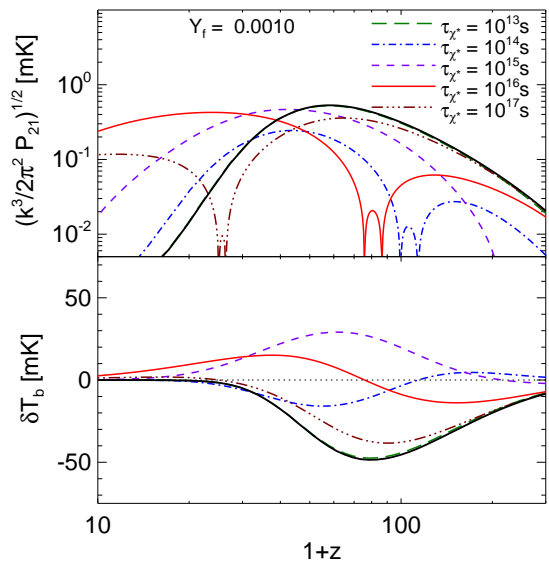}
\caption{\label{fig:igm_21cm_yf} 21 cm signals: 
 the fluctuation amplitude for an arbitrary scale, $k = 0.04\Mpc^{-1}$
 (\emph{top}), and mean (sky-averaged) signal (\emph{bottom}), for
 various values of the lifetime, $\tau_{\chi^*}$, with $Y_f$ held
 fixed. 
 The baseline scenario, with no energy injection from WIMPs,
 is shown in both panels (\emph{thick solid line}), and $\delta T_b=0$ is
 shown in the bottom panel (\emph{dotted line}).  In all cases we
 take $Y_f = 10^{-3} \epsilon_{b}$. Line styles and colors are the same
 as Fig. \ref{fig:igm_temp_yf} for easy comparison. 
}
\end{figure}
\begin{figure}
\includegraphics[width=3in]{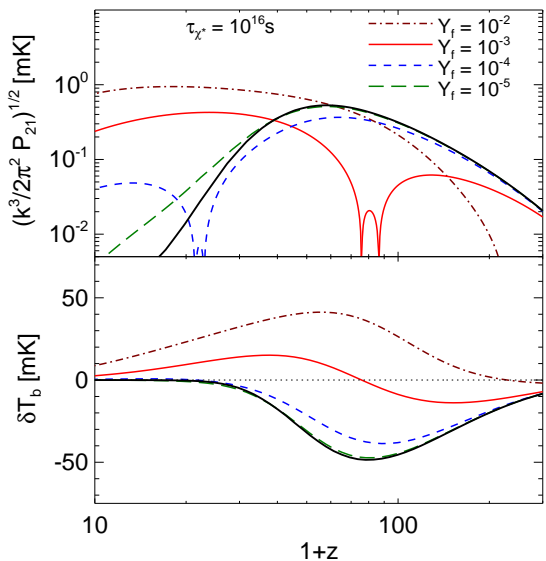}
\caption{\label{fig:igm_21cm_tau} Same as Fig. \ref{fig:igm_21cm_yf}
 but for
 various values of $Y_f$ with $\tau_{\chi^*} = 10^{16}$s.  The
 fiducial model ($Y_f=10^{-3}$ and $\tau_{\chi^*} = 10^{16}$s) is
 represented by a red solid line in both figures.
}
\end{figure}

The hyperfine (spin-flip) transition in neutral atomic hydrogen
provides a source of opacity to the cosmic microwave background from
the time the gas temperature decouples from the CMB ($z=200$) until
reionization ($z\sim10-20$).  In the standard scenario, $T_S<T_\gamma$
and the line appears in absorption.  If sufficient energy is injected
by new physics (or any other mechanism) such that $T_S>T_\gamma$, the
line will appear in emission.  Observations of this line have the
potential to constrain the evolution of the matter power spectrum, and
reveal new information about the first sources of ionizing radiation
\cite{Loeb:2003ya, Furlanetto:2006wp, Mapelli:2006ej}.
  Several projects are already underway to observe the line at
$z\sim7 - 14$ (e.g., the 
Murchison Widefield Array \footnote{
\texttt{http://www.haystack.mit.edu/ast/arrays/mwa}},
LOFAR \footnote{\texttt{http://www.lofar.org}},
and others).  The heroic observing efforts now underway may be
sensitive to the expected signal from relic XDM excitation
within 5-10 years. 

Following standard practice, we assume the spin temperature $T_S$ is
well defined in each volume element of space, and neglect the subtle
variation of $T_S$ with atomic velocity \cite{Hirata:2006bn} needed
for high-precision calculations.  We relate $T_S$ to the ratio of the
number densities of ground state and excited atoms via
\be
\frac{n_1}{n_0} = \frac{g_1}{g_0} \exp\left(\frac{-T_*}{T_S}\right)
\ee
where $g_1/g_0=3$ is the degeneracy factor, and $T_*=0.068\K$ is the
temperature corresponding to the energy of the transition. 

The mean (sky-averaged) signal $\bar{\delta T_b}$ is given by 
\be
\bar{\delta T_b}  = 27(1-x_i)\left(\frac{1+z}{10}\right)^{1/2}
\left(\frac{T_S-T_\gamma}{T_S}\right) \mK\,,
\ee
where $x_i$ is the ionization fraction, standard cosmological
parameters are assumed, and radial peculiar velocity of the gas is
neglected \cite{Furlanetto:2006wp}.  Figs. \ref{fig:igm_21cm_yf} and
\ref{fig:igm_21cm_tau} show $\bar{\delta T_b}$ for the same set of
parameters as Figs. \ref{fig:igm_temp_yf} and \ref{fig:igm_temp_tau}.

Computation of the $T_S$ history involves collisional coupling of
$T_S$ to $T_K$ via both H-H and H-$e^-$ collisions
\cite{Zygelman:2005,Furlanetto:2006wp,Furlanetto:2007}.  Another
important effect is the Wouthuysen-Field coupling, in which photons in
the Ly $\alpha$ resonance region exchange energy with atoms via
Doppler shift, and also couple to the hyperfine transition via Raman
scattering \cite{Wouthuysen:1952,Field:1959,Hirata:2005mz}.  The net
result is that the presence of Ly $\alpha$ photons more tightly
couples $T_S$ to the gas temperature.  We follow standard procedure
\cite{Loeb:2003ya,Furlanetto:2006wp} to compute $T_S$ and the gas
temperature, $T_K$ (Fig. \ref{fig:gastemp}).

Because the galactic synchrotron foreground is orders of magnitude
brighter than $\bar{\delta T_b}$, the more relevant signal to consider
is the signal from fluctuations in the gas density.  These
fluctuations cause the observed signal to vary both because the
density of the gas (and therefore optical depth) varies, and also the
ionization state and temperature.  Following \cite{Furlanetto:2006wp}
we combine these effects and present the expected amplitude at
wavenumber $k=0.04\Mpc^{-1}$ (Figs. \ref{fig:igm_21cm_yf} and
\ref{fig:igm_21cm_tau}).  This scale is arbitrary, but is chosen
to be roughly the largest scale observable by the current generation
of arrays.  It is clear that at $z\sim10$ XDM energy injection could
provide a substantial enhancement over the baseline scenario.
However, the first stars or other astrophysical sources could also
produce a dramatic signal, so mapping the history at higher $z$ would
be necessary to unambiguously identify any new physics.

\section{Conclusions}
Exciting dark matter (XDM) was invented to explain the 511 keV signal in the
center of the Milky Way. This was achieved by converting WIMP kinetic energy into
excitations, and then into $e^\pm$ pairs from subsequent decays.  An unintended feature of
XDM is that the lifetime of the excited state $\tau_\chi^*$ may be
long ($\sim 10^{13} - 10^{18}\s$).  We have computed the expected
relic excitation fraction $Y_f$ left over from the early Universe, and
explored a variety of observable consequences for various $Y_f$ values
and lifetimes.  

Such features would be generic in a wide class of models with excited states, beyond the simple model of
\cite{Finkbeiner:2007kk}. We find that for the parameters that explain the 511
keV signal (500 GeV mass, $\tilde{\sigma}_{mr} \approx 0.1-50$) the
expected $Y_f$ is small.  However, for a higher mass particle and
weaker cross section, very substantial relic densities (up to $Y_f =
10^{-2}$) are possible, which may be generic in models with multiple excited states.  The thermal, ionization and spin temperature histories of the universe are sensitive tests of the detailed physics of dark matter properties, to which collider tests may be insensitive. Upcoming probes of this era may show anomalies, giving essential insight into the nature of dark matter.
\acknowledgments
DPF is partially supported by NASA LTSA grant NAG5-12972.  NP is
supported by NASA Hubble Fellowship HST-HF-01200.01 awarded by the
Space Telescope Science Institute, which is operated by the
Association of Universities for Research in Astronomy, Inc., for NASA,
under contract NAS 5-26555.  NP is also supported by an LBNL
Chamberlain Fellowship.  This work was partially supported by the
Director, Office of Science, of the U.S.  Department of Energy under
Contract No. DE-AC02-05CH11231.  NW is supported by NSF CAREER grant
PHY-0449818 and DOE OJI grant \#DE-FG02-06E R41417.  We thank an
anonymous referee for helpful comments on the current CMB constraints.
We are grateful to the Stanford Physics Department
for hospitality at the inception of this project.

\appendix

\section{Estimating (de-)excitation rates}
\label{sec:rates}

The velocity excitation/de-excitation rate coefficients ($k_D,k_E$; 
scatterings per time per density) are given by 
\begin{equation}
k_{D,E}({\bf r}) = \int d^{3}v_{1} d^{3}v_{2} f({\bf v}_{1},{\bf r})
f({\bf v}_{2},{\bf r}) \sigma_{D,E}(v_{rel}) v_{rel} \,\,,
\end{equation}
where $f({\bf v},{\bf r})$ is the phase space density of particles with velocity ${\bf v}$ 
at position ${\bf r}$, $\sigma_{D,E}(v_{rel})$ is the inelastic scattering cross-section as a 
function of the relative velocity $v_{rel} = |{\bf v}_1 - {\bf v}_2|$. Assuming the 
universe is homogeneous and the particles are non-relativistic, the phase-space density
is given by the Maxwell-Boltzmann distribution,
\begin{equation}
f({\bf v},{\bf r}) = \left(\frac{m}{2\pi T}\right)^{3/2}\exp\left(\frac{-mv^{2}}{2T}\right)
\end{equation}
where $m$ is the mass of the particle, and $T$ is the kinetic temperature (in energy
units). It is convenient to transform to center of mass variables, ${\bf V}_{cm}$ and ${\bf v}_{rel}$
which decouples the velocity integrals over both particles,
\begin{equation}
k_{D,E} = \left(\frac{m}{4\pi T}\right)^{3/2}\int d^{3}v_{rel} 
\exp\left(\frac{-mv_{rel}^{2}}{4T}\right) \sigma(v_{rel}) v_{rel} \,\,.
\end{equation} 
If, following \cite{Finkbeiner:2007kk}, we assume the de-excitation
cross-section, $\sigma_{mr}$, is independent of velocity, we obtain
for the de-excitation rate,
\begin{equation}
\label{eq:app_kd}
k_{D} = 4\sigma_{mr} \sqrt{\frac{T}{\pi m}} \,\,.
\end{equation}
which scales as $\sqrt{T}$ as expected from dimensional arguments. We
approximate the excitation rate (Eq. 2 of \cite{Finkbeiner:2007kk}) with
\begin{eqnarray}
\sigma v_{rel} = &  \sigma_{mr} \sqrt{v_{rel}^{2} - 4 \delta/m} & \, v_{rel}^{2} \ge 4\delta/m \,\nonumber \\
= &  0 & \, v_{rel}^{2} < 4\delta/m \,,
\end{eqnarray}
where $\delta$ is the energy splitting between $\chi$ and $\chi^{*}$ and
we assume the same (velocity-independent) cross-section, $\sigma_{mr}$
as for de-excitation.  Setting $x=v_{rel}\sqrt{m/4T}$ yields
\begin{equation}
k_{E} = 2k_{D} \int_{\sqrt{\delta/T}}^{\infty} dx \, x^{2} e^{-x^2} \sqrt{x^{2}-\delta/T}  \,\,.
\end{equation}
Changing variables,  
\begin{equation}
k_{E} = k_{D} \int_{\delta/T}^{\infty} dy \, \sqrt{y} e^{-y} \sqrt{y-\delta/T} \,\,,
\end{equation}
which is a known integral (Eq. 3.383 of \cite{Gradshteyn:1994}) giving
\begin{equation}
\label{eq:app_ke}
k_{E} = k_{D} \left(\frac{\delta}{2T}\right) K_{1}\left(\frac{\delta}{2T}\right) 
    \exp\left(\frac{-\delta}{2T}\right) \,\,,
\end{equation}
where $K_{1}$ is the modified Bessel function of the second kind. Using the fact that $z K_{1}(z) \approx 1$
as $z \rightarrow 0$, we see that $k_{E} \approx k_{D}$ for $T \gg \delta$. As $T \ll \delta$, we find
\begin{equation}
\label{eq:ke_asymp1}
k_{E} \sim k_{D} \sqrt{\frac{\pi \delta}{4 T}} \exp\left(\frac{-\delta}{T}\right) \,.
\end{equation}
This implies that $k_{E}$ decreases slower than the naive Boltzmann scaling suggests, although 
the correction only grows as $\sqrt{\delta/T}$. The asymptotic expressions suggest an approximation,
\begin{equation}
\label{eq:app_ke_approx}
k_{E} \approx k_{D} \sqrt{\left(1+\frac{\pi \delta}{4 T}\right)} \exp\left(\frac{-\delta}{T}\right) \,\,;
\end{equation}  
this approximation agrees with Eq. \ref{eq:app_ke} to within a few
percent for $T>\delta$. 

Finally, it is useful to estimate a numerical value for these rate
coefficients. As in \cite{Finkbeiner:2007kk}, we assume $\sigma_{mr}$
is determined by the momentum transfer, $\sigma_{mr} = \tilde{\sigma}_{mr}/\delta m$, 
where $\tilde{\sigma}_{mr}$ is assumed to be independent of $\delta$ and $m$. 
Choosing fiducial values, this gives,
\begin{eqnarray}
\label{eq:app_kd_fid}
&& k_{D} \sim 2 \tilde{\sigma}_{mr} {\rm GeV^{-2}} \nonumber \\
&&\times \left(\frac{\delta}{1 {\rm MeV}}\right)^{-1} 
\left(\frac{T}{1 {\rm GeV}}\right)^{1/2} \left(\frac{m}{100 {\rm GeV}}\right)^{-3/2} \,\,.
\end{eqnarray}
with $1 \GeV^{-2} = 3.90\times10^{-28}\cm^2$. 

\newpage
\onecolumngrid
\bibliography{xdm_resid}
\bibliographystyle{apsrev}

\end{document}